\documentclass[prl,twocolumn,showpacs,amsfonts,amsmath,floatfix]{revtex4}
\usepackage{graphicx}
\usepackage{epsfig}
\usepackage{epstopdf}
\newcommand{\Journal}[4]{#1 {\bf #2}, #3 (#4)}
\newcommand{\PRL}{Phys. Rev. Lett.}
\newcommand{\PRA}{Phys. Rev. A}

\newcommand{\lt}{\left(}
\newcommand{\rt}{\right)}
\newcommand{\lqu}{\left[}
\newcommand{\rqu}{\right]}

\newcommand{\be}{\begin{equation}}
\newcommand{\ee}{\end{equation}}
\newcommand{\ba}{\begin{eqnarray}}
\newcommand{\ea}{\end{eqnarray}}
\newcommand{\fr}{\frac}
\newcommand{\nn}{\nonumber}
\begin{document}

\title{Optimal control of quantum superpositions in a bosonic Josephson junction}

\author{M. Lapert$^1$}
\author{G. Ferrini$^{2,3}$}
\author{D. Sugny$^1$}
\email{dominique.sugny@u-bourgogne.fr} \affiliation{$^1$
Laboratoire
  Interdisciplinaire Carnot de Bourgogne (ICB), UMR 5209 CNRS-Universit\'e de
  Bourgogne, 9 Av. A. Savary, BP 47 870, F-21078 DIJON Cedex, FRANCE}
  \affiliation{$^2$ Laboratoire de Physique et Mod\'elisation des Milieux Condens\'es,
  Universit\'e Joseph Fourier and CNRS, Bo\^ite Postale 166, F-38042 Grenoble, France}
  \affiliation{$^3$ Laboratoire Kastler Brossel, Universit\'e Pierre et Marie Curie-Paris 6, ENS, CNRS; 4 place Jussieu, 75252 Paris, France}
\date{\today}

\begin{abstract}
We show how to optimally control the creation of quantum
superpositions in a bosonic Josephson junction within the two-site
Bose-Hubbard model framework. Both geometric and purely numerical
optimal control approaches are used, the former providing a
generalization of  the proposal of Micheli \emph{et al} [Phys.
Rev. A \textbf{67}, 013607 (2003)]. While this method is shown not
to lead to significant improvements in terms of time of formation
and fidelity of the superposition,  a numerical  optimal control
approach appears more promising, as it allows to create an almost
perfect superposition, within a time short compared to other
existing protocols. We analyze the robustness of the optimal
solution against atom number variations.
Finally, we discuss to which extent these optimal solutions could
be implemented with the state of art technology.
\end{abstract}

 \maketitle

\section{Introduction}

The high controllability of the experimental parameters of ultracold atomic systems renders them versatile candidates for the implemention of quantum information protocols.
In particular, the ability of
tailoring traps of various geometry~\cite{Lesanovsky,Fortagh} and
of tuning the interatomic interactions~\cite{Julienne} makes
ultracold atomic systems a unique playground for applications in
quantum technologies. Recently, the Bosonic Josephson Junction
(BJJ), a system of two-mode ultracold bosons, has received a large
theoretical~\cite{Smerzi97,Micheli,Smerzi09} and
experimental~\cite{Oberthaler05,Oberthaler08,Oberthaler10,Treutlein10}
interest. It has been proposed~\cite{Kitagawa93, Sorensen} and
experimentally
demonstrated~\cite{Oberthaler08,Oberthaler10,Treutlein10} that
such a system allows for the creation of atomic squeezed states.
These non classical states can be used in an interferometric
protocol to improve the phase sensitivity reducing it below the
shot-noise limit $1/\sqrt{N}$ ~\cite{Wineland}, as recently
demonstrated in BJJs by C. Gross {\it et al} \cite{Oberthaler10}.
A further enhancement of precision in atom-interferometry has been
predicted to be reached by the use of macroscopic superpositions
of atomic coherent states. Such states allow in principle for a
scaling of the phase uncertainty as $1/N$, which corresponds to the ``Heisenberg
limit"~\cite{Giovannetti, Smerzi09}. Their experimental
realization would be hence of primary importance both for
applications in metrology, and for the fundamental investigation
of the boundary between classical and quantum dynamics~\cite{Haroche_Book}.

These states however have so far been realized rather with a
mesoscopic number of particles up to few tens in systems of
ions~\cite{Leibfried}, nuclear spins \cite{Jones} or photons in a cavity~\cite{Deleglise}. Several proposals to create macroscopic quantum superpositions in BJJs have been advanced \cite{Micheli,noi,Piazza_08,Dunningham,Chr_preprint_bifurcation}.
The realization of superposition states with a macroscopic number of
particles is challenging because of their fragility with respect
to decoherence, induced by particle losses~\cite{Sinatra}, phase noise~\cite{Ferrini_10},
collisions with thermal atoms~\cite{Anglin,Witthaut}, interaction
with the electromagnetic field~\cite{Moore06}, and random
fluctuations of the trapping potential~\cite{Khodorkovsky}.
Therefore, it is of great interest to provide a protocol to create
macroscopic quantum superpositions in the shortest time possible,
i.e. before that decoherence becomes effective. The optimization
of the fidelity with which such states can be formed in a given
time, this latter being chosen as small as possible, represents
another important task. Some intuitive techniques have been
developed over the past few years to fulfill such objectives
\cite{Micheli,noi,Piazza_08,Dunningham,watanabe,mahmud}. However,
these approaches are not able to reach the physical limits of the
best possible performance in terms of duration and efficiency.
These limits can be established by using optimal control theory
\cite{bryson,geometry,brif}. 
Optimization techniques have already been applied in the context
of ultracold atoms (see \cite{grond1,kosloff1,kosloff2,doria} to
cite a few) to improve, e.g., the performances of an
interferometer \cite{grond2}. In this paper we apply such a method
to provide a protocol for the efficient creation of macroscopic
superpositions in BJJs. In particular, we show how an almost
perfect macroscopic superposition can be created in a relatively
short time by means of a numeric optimal control approach.

Before doing this, it is instructive to consider a geometric optimal control approach \cite{geometry,boscainbook}.  This method provides a generalization of the proposal of Micheli \emph{et al} reported in Ref.~\cite{Micheli}, and allows to obtain an estimation for a bound on the time of formation of macroscopic superpositions in the bosonic Josephson junction.
In a summarized way, geometric optimal control is a vast domain where
the optimal control problems are solved by using tools of geometry
and Hamiltonian dynamics. Surprisingly, while optimal control is
known in quantum mechanics since the eighties, only few results of
geometric control have been published up to date
\cite{boscain1,khaneja}. Recent advances in this domain have permitted to attack problems of increasing difficulty. In
particular, some of the authors have applied this method with
success in fundamental quantum control problems \cite{sugnyfund}
and in Nuclear Magnetic Resonance \cite{NMR}. Due to its geometric
framework, this method is however intrinsically limited to systems
with few degrees of freedom. In our example, we describe the BJJ in the two-mode approximation by the two-site Bose-Hubbard model, and we
consider its semi-classical limit,
which is valid when the number of particles is large. 
In this limit, the system can be described in terms of two classical conjugated variables, which can be related to the polar and azimuthal angles of a Bloch sphere of radius $N/2$, where $N$ is the total number of atoms. Hence the tools of geometric optimal control theory can be applied. Within this approximation, we
show how to reach in minimum time a quantum superposition state. 
Due to the semi-classical approximation, the efficiency of the optimal
solution is however limited in the original quantum domain.

We then determine the solution of the initial quantum problem by using a
purely numerical approach, the monotonic convergent algorithm. The
efficiency and the flexibility of such algorithms have been
demonstrated in a large number of studies \cite{tannorbook,mono}.
In order to guide the numerical optimization, we use a coupling
between the geometric and the numerical approaches in the sense
that the geometric solution is used as a solution to initialize the
numerical algorithm. 
The duration of the numerical solution necessary to produce with efficiency a superposition state is shown to be about ten times larger than with the geometric approach. We analyze the robustness of this factor with respect to the number of particles.

The paper is organized as follows. We introduce in Sec. \ref{sec2}
the model system, i.e. the bosonic Josephson junction in the
two-mode approximation. We also recall how to construct the
corresponding classical limit when the particle number is large,
and we set the control problem under study. We then briefly recall
how to produce macroscopic superpositions within the protocol of
Ref.~\cite{Micheli}. Section \ref{sec4} focuses on the numerical
results of the geometric and purely numerical methods, a
comparison with the existing results is also made. Finally, we
analyze the robustness of the optimal solution against atom number
fluctuations. In Sec. \ref{sec-exp}, we discuss the possible
experimental implementations of our optimal solutions with the
state-of-the-art technology
\cite{Oberthaler08,Oberthaler10,Gross_thesis}. Conclusion and
prospective views are given in a final section. Section \ref{sec3}
of the appendix is devoted to the presentation of the geometric
and numerical optimal control approaches. Some technical
computations are also reported in Sec. \ref{sec_tests} and
\ref{ana} of the appendix.

\section{The bosonic Josephson junction and the setting of the problem}

\subsection{The model system}\label{sec2}

The Bosonic Josephson Junction can be described in the two-mode
approximation (assuming fixed the total number of particles $N$)
by the spin-Hamiltonian
\begin{equation}
\label{eq:ham_dopo_mapping} \hat{H}= \chi \hat J_z^2 + \delta \hat
J_z - \Omega \hat J_x\;,
\end{equation}
where the angular momentum operators $\hat J_x$, $\hat J_y$, and
$\hat J_z$ are related to the annihilation operator $\hat{a}_j$ of
an atom in the mode $j=1,2$ respectively by $\hat J_x=(\hat
a^\dagger_1 \hat a_2+ \hat a^\dagger_2 \hat a_1)/2$, $\hat J_y=-i
(\hat a^\dagger_1 \hat a_2-\hat a^\dagger_2 \hat a_1)/2$, and
$\hat J_z \equiv \hat n = (\hat a^\dagger_1 \hat a_1- \hat
a^\dagger_2 \hat a_2)/2$ (number imbalance
operator)~\cite{Milburn,notazioni}.
This model describes both a system of cold atoms in two distinct hyperfine states, trapped in a harmonic potential (internal BJJ)~\cite{Hall98,Oberthaler10}, and a system of bosons confined in a double well potential (external BJJ)~\cite{Oberthaler08}. The interaction constant, $\chi=U_1+U_2 - 2 U_{12}$, related to the atom-atom interaction energy $U_i$ in the mode $i$ and to the cross-interaction $U_{12}$, can be tuned to some extent, by exploiting e.g. Feschbach resonances~\cite{Oberthaler10}. The cross-interaction term $U_{12}$ is usually taken to be zero in the external BJJ, because of the spatial separation of the two modes. The parameter $\delta$ is related to the energy
imbalance between the two modes and to the interaction imbalance, and is defined by $\delta = (E_1 - E_2) + (N-1) (U_1 - U_2)/2$. In the external case the Rabi coupling $\Omega$, which represents the tunneling term, is usually positive, i.e. $\Omega > 0$, and can be tuned by changing the height of the barrier separating the two wells. A negative coupling $\Omega$ can be engineered by applying a drive to the barrier height~\cite{Arimondo}. Such a coupling is however far more efficiently controllable in the internal setup, where by modulating amplitude and phase of resonant microwave and radiofrequency fields shone on the atoms, both amplitude and sign of $\Omega$ can be tuned instantaneously with respect to the other time scales of the problem \cite{Oberthaler10} (see also the concluding section for a discussion about the experiments). Hence, to develop our control protocol we will keep fixed the parameter $\chi$ with $\delta=0$ and use $\Omega$ as control field, having in mind the internal BJJ setup which appears more suitable for the experimental implementation of the protocol.

As initial state of the dynamical proposal described in the following we will take an atomic coherent state where all of the atoms occupy the same
one-particle state~\cite{husimi}
\begin{equation} \label{eq-coherent_state}
|\theta, \phi \rangle=\sum_{n=-N/2}^{N/2} \left(  \begin{array} {c} N \\
 n+\frac{N}{2} \end{array}\right)^{1/2} \frac{\alpha^{n+N/2}}{(1+|\alpha|^2)^{N/2}}\, |n\rangle,
\end{equation}
with $\alpha \equiv \tan(\theta/2) \exp(-i\phi)$, and $| n
\rangle$ the Fock state satisfying
\begin{equation}
\label{eq:fock_states}
\hat J_z |n\rangle = n | n\rangle.
\end{equation}
Each coherent state (\ref{eq-coherent_state}) can be represented
on the Bloch sphere as a circle, the center of which has
coordinates given by the expectation value of the angular momentum
operator $\langle \theta, \phi | \vec{J}| \theta, \phi \rangle  =
N (\sin \theta \cos \phi ,\sin \theta \sin \phi , -\cos
\theta)/2$. Since the quantum fluctuations of the angular momentum
operators in each direction tangential to the Bloch sphere in the
point $\langle \theta, \phi | \vec{J}| \theta, \phi \rangle$ are
given by $\sqrt{(\Delta J_i)^2} = \sqrt{N/4}$, as an order of
magnitude for the radius of the circle we can take $\sigma =
\sqrt{N}$.
Starting from the Heisenberg
equations for $\vec{J}$, one obtains by using a semi-classical limit \cite{Micheli}
that the angular coordinates $(\theta,\phi)$ satisfy the system
\begin{equation}
\left(  \begin{array} {c} \dot{\theta} \\
 \dot{\phi} \end{array}\right)=
\left(  \begin{array} {c} -\Omega\sin\phi \\
 \delta-N\chi\cos\theta-\Omega\cot\theta\cos\phi \end{array}\right)
\end{equation}
which can be written as follows
\begin{equation}\label{eqsph}
\left(  \begin{array} {c} \dot{\theta} \\
 \dot{\phi} \end{array}\right)=\frac{\chi N}{2}
 \left(  \begin{array} {c} -\omega\sin\phi \\
 \Delta-2\cos\theta-\omega\cot\theta\cos\phi \end{array}\right)
\end{equation}
by introducing the parameters $\Delta=2\delta/(\chi N)$ and
$\omega=2\Omega/(\chi N)$. In the following, we consider the
situation where $\delta=0$ and the scalar factor $\chi N/2$ is
assumed to be constant. The system is then controlled by only one
parameter, namely $\omega$.  
Equation (\ref{eqsph}) can also be written in a more
compact form as
\begin{equation}
\left(  \begin{array} {c} \dot{\theta} \\
 \dot{\phi} \end{array}\right)=\vec{F}+\omega\vec{G}
\end{equation}
where $\vec{F}$ and $\vec{G}$ are two vector fields of coordinates
$\frac{\chi N}{2}(0,-2\cos\theta)$ and $\frac{\chi
N}{2}(-\sin\phi,-\cot\theta\cos\phi)$.

The dynamics generated by the Hamiltonian
(\ref{eq:ham_dopo_mapping}) can produce macroscopic
superpositions of coherent states at specific
times~\cite{Micheli,Piazza_08,noi}. In order to create such
superpositions in an optimized way, we formulate this problem as
an optimal control problem. We first assume that the system is
initially in the state $|\pi/2,\pi\rangle$, the goal being to design
a ``field" $\omega(t)$ (we follow here the control terminology)
such that the system reaches a superposition state. 
We will consider as target states perfect macroscopic
superpositions of coherent states (\ref{eq-coherent_state}),
commonly referred to as "Schr\"odinger's cat state". For example,
the state \be \label{cat1} |\textrm{Cat}_1\rangle=
\fr{1}{\sqrt{2}} (|\theta = 0\rangle +|\theta = \pi\rangle) \ee is
the superposition of the two coherent states on the poles of the
Bloch sphere. This state is also know as a "NOON cat state",
because of its equivalent expression on the basis of the mode
occupation $|\textrm{Cat}_1\rangle = \frac{1}{\sqrt{2}} (|N,0
\rangle + |0,N \rangle)$. The rotation of the previous state by
$\pi/2$ around the $y$ axis leads to the "phase cat state"
\be
\label{cat2}
|\textrm{Cat}_2\rangle= \fr{1}{\sqrt{2}} (|\theta = \pi/2, \phi = 0 \rangle +|\theta = \pi/2, \phi = \pi \rangle),
\ee
i.e. the superposition of the two coherent states located on the equator of the Bloch sphere along the $x$- axis.
Among all the functions $\omega(t)$ allowing to reach such targets, the
optimal solution will be the one minimizing a
given cost, such as the duration of the control (with
a bound on the amplitude of the control field) or the energy of the field (with a fixed
control duration). In this framework, optimal control theory can be
viewed as a vast machinery aiming at solving such problems either
analytically (or with a very high numerical precision) for
geometric methods, or numerically for the optimal control
algorithms.

\subsection{Creation of macroscopic superposition in the proposal of Micheli \emph{et al}}\label{sec2bis}

In Ref.~\cite{Micheli}, a proposal for the creation of a NOON state has been given, based on a semi-classical argument. We briefly recall here this method, in the direction of which the experiments of Ref.~\cite{Chr_preprint_bifurcation} are performed. 

In the classical model, the dynamics can be visualized by means of trajectories on the Bloch sphere parameterized by the coordinates $\theta$ and $\phi$ (see Fig.~\ref{fig1}, and also Refs.~\cite{Micheli,Smerzi97,Chr_preprint_bifurcation}).
For $\omega= 2$ a bifurcation occurs in the model, and the fixed points of the system change: in the Rabi regime ($\omega > 2$) $F_0 = (\theta = \pi/2,\phi = 0)$ and $F_\pi = (\theta = \pi/2, \phi = \pi)$ are the two stable fixed points, while in the Josephson regime ($\omega < 2$) $F_\pi$ becomes unstable and two new stable fixed points at $F_{\pm} = (\cos \theta = \mp \sqrt{1 - (\omega/2)^2} , \phi = \pi)$ appear \cite{Smerzi97,Chr_preprint_bifurcation}. In the latter regime, trajectories in which the number imbalance can not be reduced to zero are allowed (macroscopic quantum self-trapping).
For each value $\omega < 2$, there exists hence a special eight-shaped trajectory which separates the macroscopic quantum self-trapping trajectories from the oscillations in which the number can take the zero value. This  special trajectory, which passes through $F_\pi$, is called the \emph{separatrix} and can be used to produce a quantum macroscopic superposition state. In particular, for $\omega = 1$ the two poles of the Bloch sphere belong to the separatrix.
The initial state of the quantum dynamics is a coherent state
centered in $F_\pi$. As recalled in Sec.~\ref{sec2}, the width
associated to its fluctuations is $\sigma = \sqrt{N}$. In a
semi-classical picture, such a state can be viewed as a cloud of
points, a half of them being in the upper hemisphere of the Bloch
sphere and a half in the bottom one, each of them evolving
according to the classical trajectories. As a result, for $\omega
= 1$ the initial wave packet evolves along the separatrix by
splitting into two outgoing parts, one going to the north pole and
the other to the south one. From a control point of view, the
solution of Ref.~\cite{Micheli} consists of choosing a field of
constant amplitude, i.e. a bang $\omega=1$. The time that it takes
for a point initially in $(\theta = \pi/2,\phi = \pi)$ to travel
along the separatrix and reach one of the poles of the Bloch
sphere is \cite{Micheli} \be \label{eq:time_cat} \chi T_c \simeq
\frac{\ln (8N)}{N}, \ee which can be taken as an estimation of the
time of formation of the macroscopic superposition. Note that this
time is much smaller than the time $T_c^{'}$ required to form a
macroscopic superposition of phase states by the \emph{quenched
dynamics}  of the BJJ (i.e. the dynamics driven by the interatomic
interactions only) which is given by $\chi T^{'}_c = \pi/2$
\cite{noi}, providing therefore a speed up with respect to the use
of that protocol. One may then check the efficiency of the
semi-classical process presented above by performing a quantum
calculation of the time evolution of the system under the
Hamiltonian (\ref{eq:ham_dopo_mapping}) with $\omega = 1$ and
$\delta = 0$. We compute the projection of the state at time $T_c$
on the cat state given in Eq.~(\ref{cat1}) \cite{Micheli} as
introduced in Appendix \ref{sec_tests}. Such a calculation leads
to a projection of $P_1 = 0.1394 $, with the parameters $\chi=1$,
$N=300$.

\section{Controlled creation of macroscopic superpositions}\label{sec4}
\subsection{Control of the semi-classical model}\label{sec4-1}

We propose to revisit the analysis presented in the previous
section in the framework of geometric optimal control theory. In
this section, we apply this approach, i.e. the Pontryagin Maximum
Principle (PMP)  and the corresponding numerical tools (see Sec.
\ref{sec3} of the appendix for a presentation of this approach) on
the semi-classical model on the sphere whose dynamics is governed
by Eq. (\ref{eqsph}). The control problem consists of creating a
Cat state in minimum time with a field bounded by the value $m$.
The role of this bound will be also analyzed in this paragraph. We
denote by $T_{min}$ the minimum time required to reach the target
state.

\begin{figure}
\centering
\includegraphics[width=2.5in]{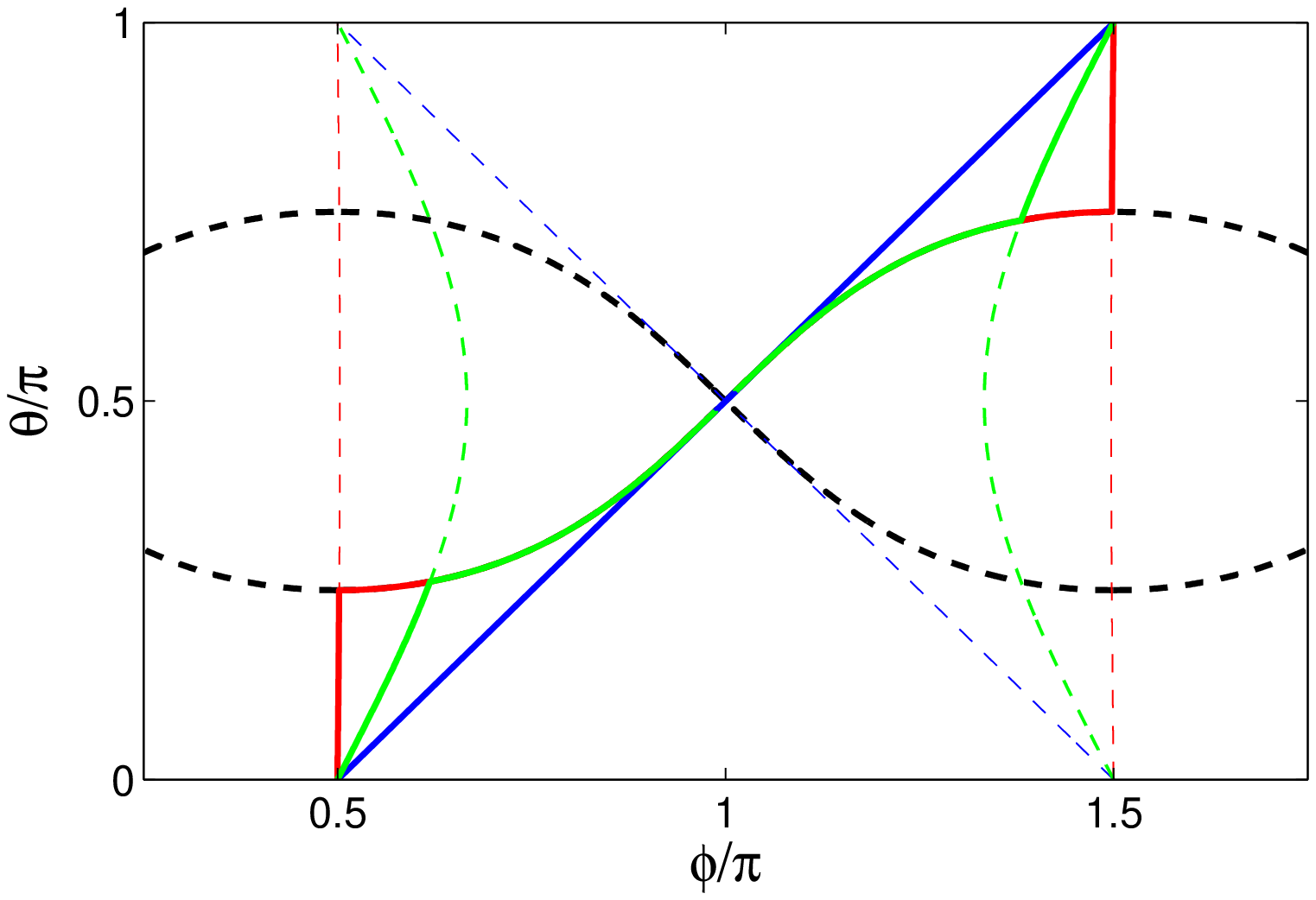}
\includegraphics[width=2.5in]{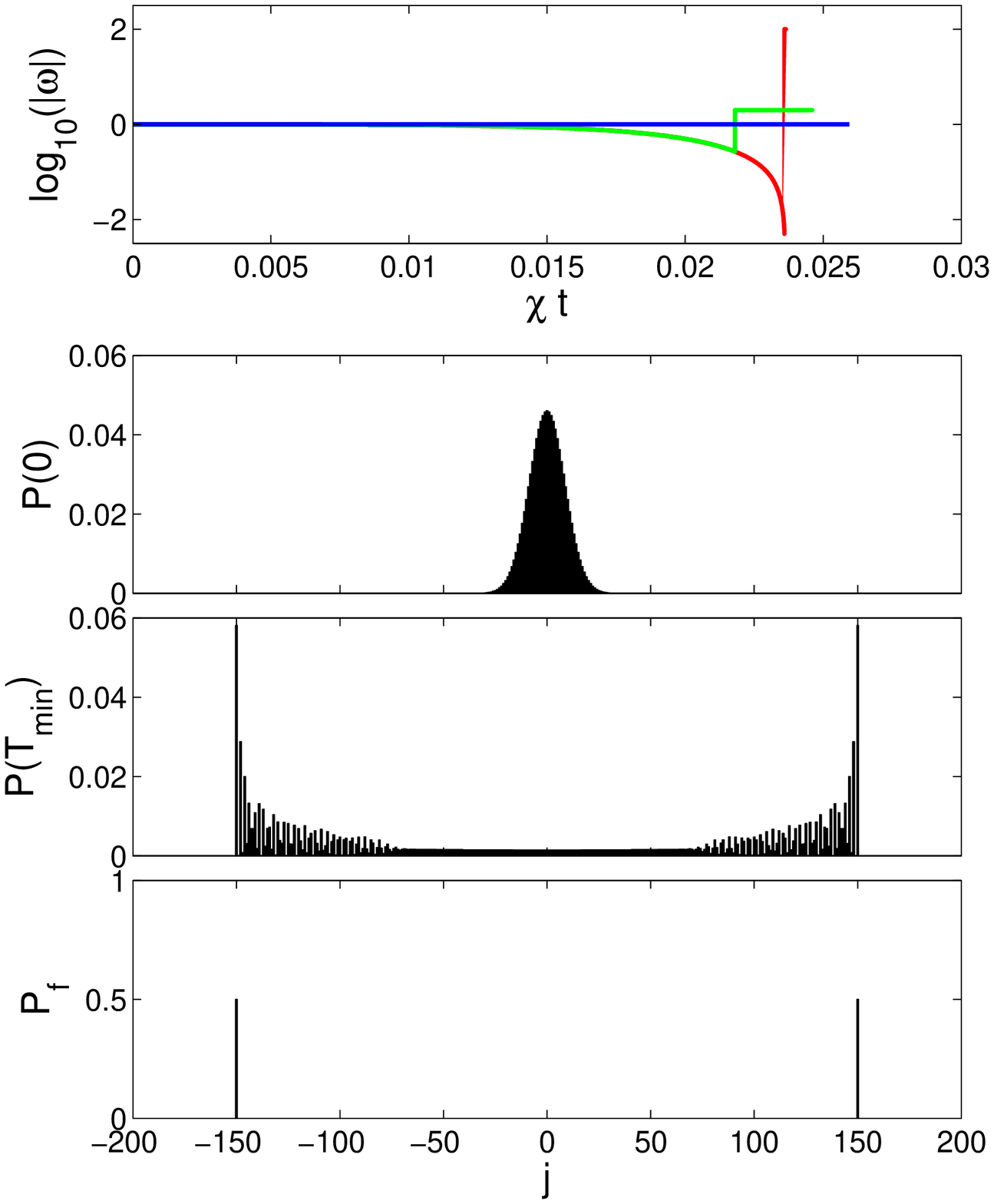}
\caption{\label{fig1} (Color online) (left) Plot in the $(\theta,\phi)$ plane of
the optimal trajectory in the semi-classical model for the bounds $m=1$, $m=2$ and
$m=100$ in blue (dark), green (light gray) and red (dark gray) respectively. The target state is the cat state $|\textrm{Cat}_1\rangle$. The dashed line indicates the position of the singular set. The dashed blue, red and green lines represent the position of the separatrix for the different bounds. (top right) Evolution of the corresponding control fields $\omega$ as a function of the dimensionless time $\chi t$. (bottom right) Fock states distribution $P(n)=|\langle
n |\psi(t)\rangle|^2$  as a function of $n$ for $t=0$, $t=T_{min}$
and the target state $|\textrm{Cat}_1\rangle$. In the quantum
calculation, the parameter $N$ is taken to be $N=300$.}
\end{figure}

We consider as initial classical
state a point on the Bloch sphere at a distance $\sigma$ of
($\theta=\pi/2$, $\phi=\pi$), corresponding to the extremum point on the uncertainty circle of the coherent state $|\theta=\pi/2, \phi=\pi \rangle$. Without loss of generality, we can
choose this point in the upper hemisphere. With this reduction,
the semi-classical control problem becomes of dimension two and all the tools presented in Sec. \ref{secgeo} of the appendix can be used. The application
of the PMP leads in the spherical coordinates $(\theta,\phi)$ to
the pseudo-Hamiltonian $H$, analogous to Eq.~(\ref{eqham})
$$
H=\vec{p}\cdot (\vec{F}+\omega\vec{G}).
$$
Introducing the components $(p_\theta,p_\phi)$ of $\vec{p}$, the
Hamiltonian $H$ reads as
$$
H=-2\cos\theta p_\phi-\omega p_\theta\sin\phi-\omega
p_\phi\cot\theta\cos\phi.
$$
Straightforward calculations following the line presented in Sec.~\ref{sec3} of the Appendix lead to
$\textrm{det}[\vec{F},\vec{G}]=-2\cos\theta\sin\phi$ and
$\textrm{det}[\vec{G},[\vec{F},\vec{G}]]=2\sin^2\phi
(1-\cos^2\theta)-2\cos^2\theta$. The singular set $S$ depicted in
Fig. \ref{fig1} is therefore given by
\begin{equation}
\phi=\arcsin [\pm \cot\theta],
\end{equation}
while the singular control field is equal to
\begin{equation}\label{eqsing}
\omega_s=\sin\theta\cos\phi.
\end{equation}
At this point, we can compute the optimal sequence to reach in
minimum time the north pole of the Bloch sphere. By symmetry of
the dynamical equations, the point of the lower hemisphere
symmetric with respect to $S$ of the initial state will reach
simultaneously the south pole. This classical simultaneous control
leads in the quantum domain to the creation of the superposition
state $|\textrm{Cat}_1\rangle$. In the optimal control theory
framework, note that there is no restriction on the value of the
bound $m$ of the control field but simple solutions can only be
obtained if the bound satisfies $m\geq 1$. We will focus on this
case in the numerical examples.

Our control protocol can be generalized to create a phase cat
$|\textrm{Cat}_2 \rangle$ in addition to the state
$|\textrm{Cat}_1 \rangle$, which is not possible in the original
proposal of Ref.~\cite{Micheli}.


The results of the time-optimal control problem for reaching both
states with various bounds show that no significant improvement in
terms of speed-up nor fidelity is achieved, compared to the
protocol of Ref.~\cite{Micheli}. Table \ref{table1} summarizes the
numerical results, while a detailed analysis is reported here
below.

\begin{center}
\begin{table}[htbp]
\begin{tabular}{ccccccc}
            \hline
            &\multicolumn{3}{c}{NOON}&\multicolumn{3}{c}{PHASE}\\
            \hline
            m & 1 & 2 & 100 & 1 & 2 & 100\\
            \hline
            $\chi t$($\times10^{-3}$) & 25.9 & 24.6 & 23.6 & 25.5 & 24.6 & 23.6\\
            $P$ & 0.139 & 0.122 & 0.116 & 0.091 & 0.100 & 0.116\\
            $F_Q/N^2$ & 0.636 & 0.596 & 0.587 & 0.514 & 0.547 & 0.586\\
            \hline
\end{tabular}
\caption{Numerical results of the semi-classical control protocol
for three different bounds, $m=1$, 2 and 100. The control duration
($\chi t$), the projection ($P$) and the Fisher information
($F_Q/N^2$) are given for the two cat states
$|\textrm{Cat}_1\rangle$ (NOON state) and $|\textrm{Cat}_2\rangle$
(Phase state) considered in this work. In the quantum calculation,
the parameter $N$ is taken to be $N=300$\label{table1}}
\end{table}
\end{center}

We first begin the analysis by the NOON state
$|\textrm{Cat}_1\rangle$. Different optimal trajectories are
displayed in Fig. \ref{fig1} for three bounds, namely
$m=1,~2,~100$. In the first situation, for $m=1$, we recover the solution of Ref. \cite{Micheli} which is only composed of a bang pulse, since the pole of the Bloch sphere can be reached by following the separatrix. More
complicated solutions can be constructed when the bound $m$ takes
larger values. In this case, one can follow the singular line from
the initial point of the dynamics to the intersection point between the separatrix and the singular locus. To simplify the discussion, we assume here that this initial point belongs to the singular set. We then leave the singular set by
using a bang pulse to follow the separatrix and to reach a pole as can be seen in Fig. \ref{fig1}. When $m\gg 1$, bang
trajectories are very close to the meridians of the Bloch sphere.
A comparison between the optimal solutions for $m=2$ and $m=100$
and the solution of Ref.~\cite{Micheli} is displayed in Fig. \ref{fig1}. With this solution, we reach the target in a time
$\chi T_{min}=0.0236$ for $m=100$ while, from Eq. (\ref{eq:time_cat}), $\chi T_c=0.0259$ in the same
conditions for $m=1$. The corresponding projections on the state $|\textrm{Cat}_1\rangle$, obtained by calculating numerically the time-evolution of the quantum state under the field $\omega(t)$, are equal to $P_1 = 0.116$ and
$P_1 = 0.139$. The values of the Fisher information are $F_Q/N^2=0.539$ and $F_Q/N^2=0.636$ respectively, which demonstrates that correlations among the component of the superposition exist, since $F_Q\simeq N^2/2\gg N$ for $N=300$. The definitions of these quantities are recalled in Appendix \ref{sec_tests}. 
The probability distribution of the Fock states is shown in Fig.~\ref{fig1}. As can be expected, such a distribution for the superposition created with our protocol mainly presents two peaks, close to the one of a perfect NOON state (see Appendix \ref{sec_tests} for details). 

\begin{figure}
\centering
\includegraphics[width=2.5in]{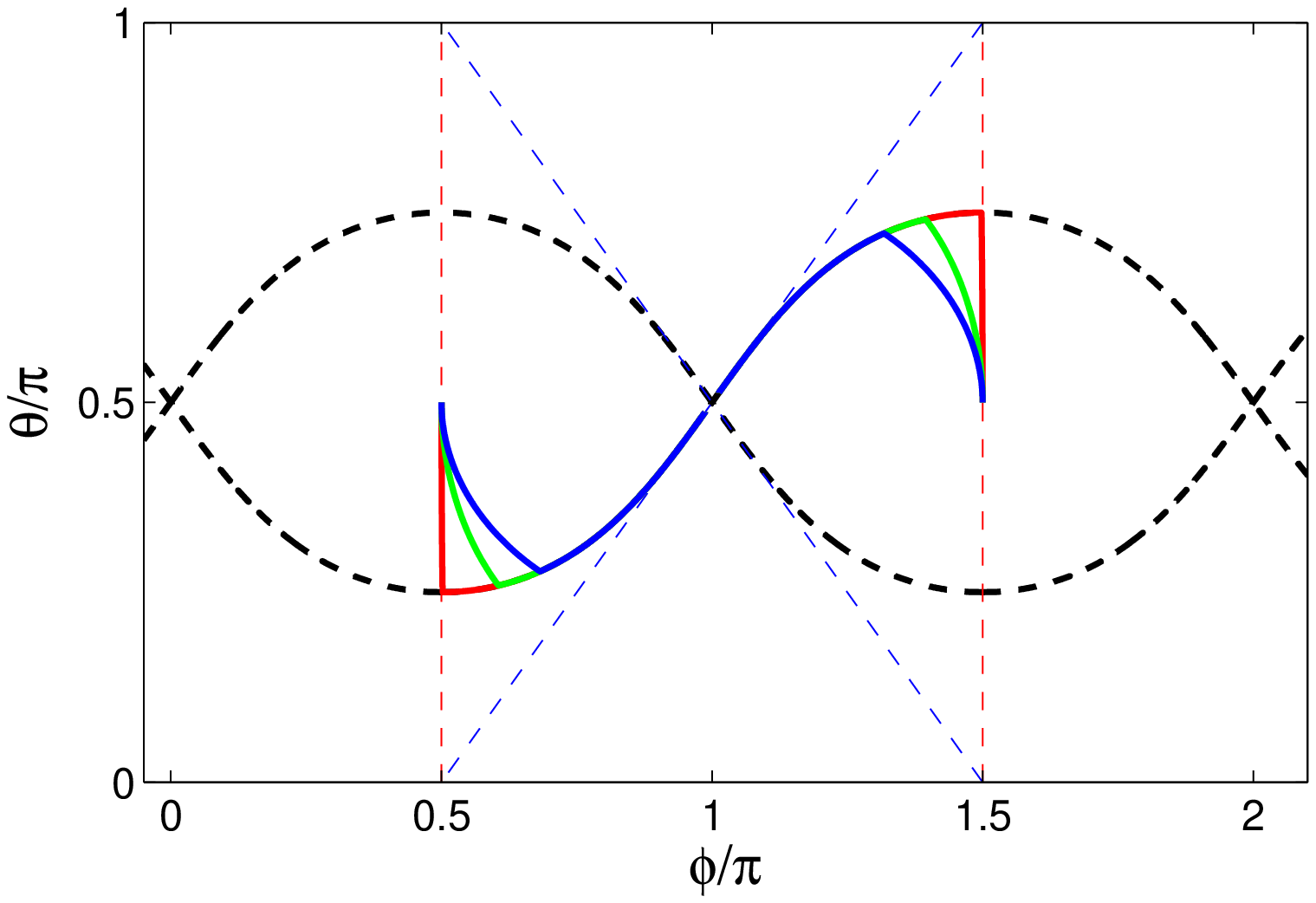}
\includegraphics[width=2.5in]{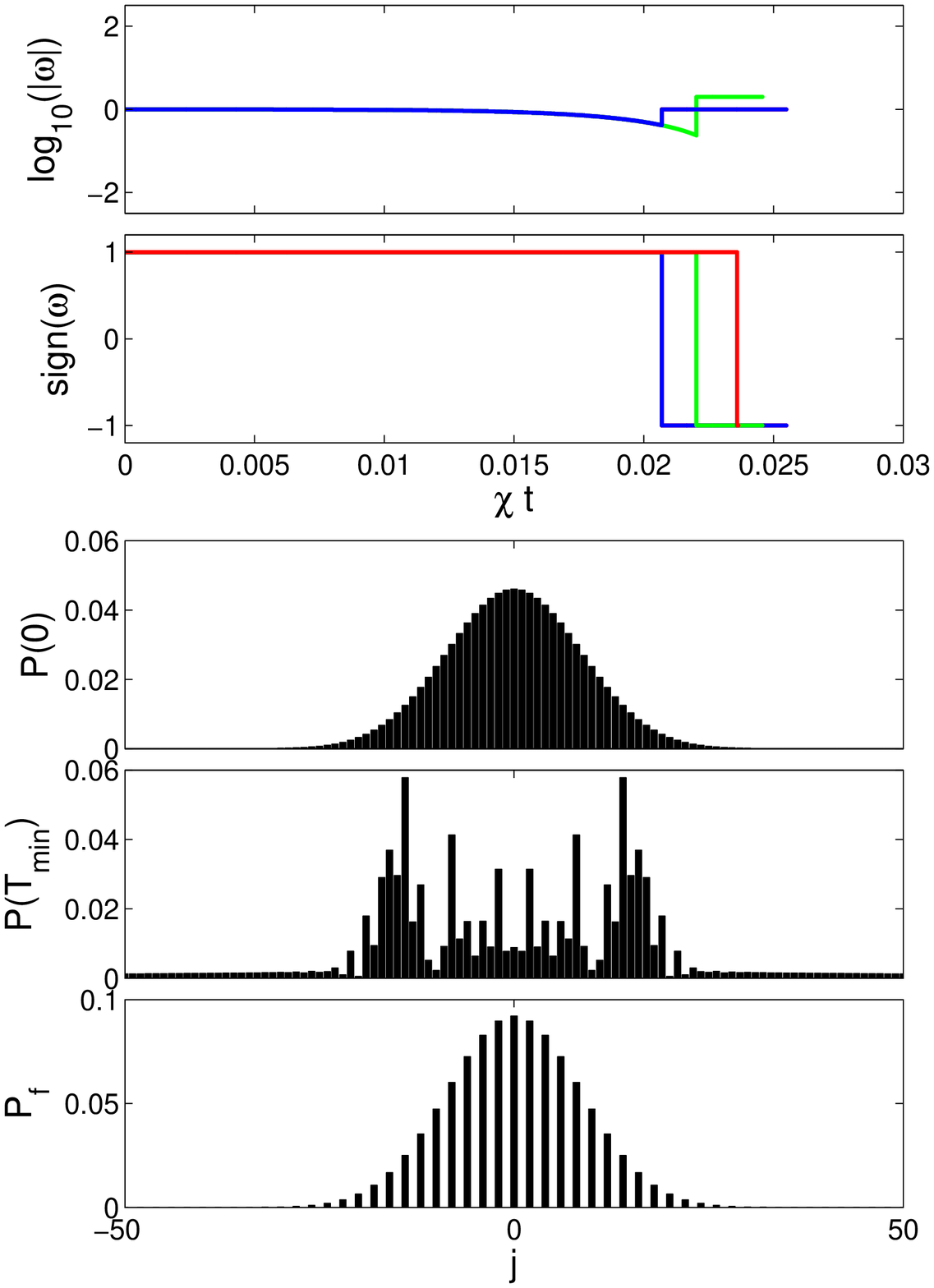}
\caption{\label{fig2} (Color online) Same as Fig. \ref{fig1} but for the target
state $|\textrm{Cat}_2\rangle$.}
\end{figure}

The same arguments can be used to describe the optimal trajectories reaching the state $|\textrm{Cat}_2\rangle$. Figure \ref{fig2} shows that the structure of the solutions are very similar for the different values of the bound $m$. The dynamics follows the singular line up to the separatrix, where a bang pulse is used to reach a point of the equator. Here, note that the sign of the bang pulse is different from the one of the first example of Fig. \ref{fig1}. Numerical results comparable with the state $|\textrm{Cat}_1\rangle$ are obtained in this case, as can be seen in Table \ref{table1}. The numerical results for the projection on the target state and the Fisher information as a function of time are reported in Fig. \ref{fig2}. Fringes appear in the Fock state distribution of the final state as explained in Appendix \ref{sec_tests}.

The time $T_{min}$ can be estimated analytically for any value of $m$ by integrating
the classical equations along the singular line. For $m\to +\infty$, we have
\begin{equation}
\label{eq:t_min_sol_geo}
T_{min}=\frac{2}{\chi
N}\int_{\phi(0)}^{\pi/2}\frac{dx}{\sqrt{\sin^2 x (1+\sin^2 x)}},
\end{equation}
where $\phi(0)$ is the coordinate of the initial point on the
singular set. Hence $T_{min}$ is inversely proportional to the total number of particles as in the case of solution of Ref.~\cite{Micheli}, from which Eq.~(\ref{eq:t_min_sol_geo}) differs by a numerical factor. The computation leading to Eq.~(\ref{eq:t_min_sol_geo}), as well as a numerical comparison between $T_c$ and $T_{min}$, are detailed in Appendix \ref{ana}.

To conclude, we stress that in these different
time-optimal-computations both the minimum time of formation of
macroscopic superpositions and the respective fidelities are
comparable to the ones obtained with the method of
Ref.~\cite{Micheli} (the numerical comparisons are summarized in
Table \ref{table1}). This shows the efficiency of the solution of
Micheli \emph{et al} for controlling the semi-classical model, and
suggests that a time of order of $T_c$ is the physical bound on
the time of formation of a macroscopic superposition in a BJJ.
Note that this non trivial result has been derived from the tools
of geometric optimal control theory. To get better results in
terms of fidelity, another approach able to tackle the fully
quantum character of the problem has to be used, which we will
address in the following section.

\subsection{Control of the quantum model}\label{sec4-2}

We apply in this paragraph the monotonic algorithm presented in
Sec. \ref{monoto} of the Appendix to maximize the projection onto the target state
at time $t=T$. We will
consider different control durations, namely $T = T_c$, $T = 5\times T_c$ and
$T = 10\times T_c$, which are multiple of the minimum time $T_c$ for a
bound $m=1$. These different tests will allow us to determine the time required to produce a cat state with efficiency.

\begin{figure}[htbp]
\includegraphics[scale=0.4]{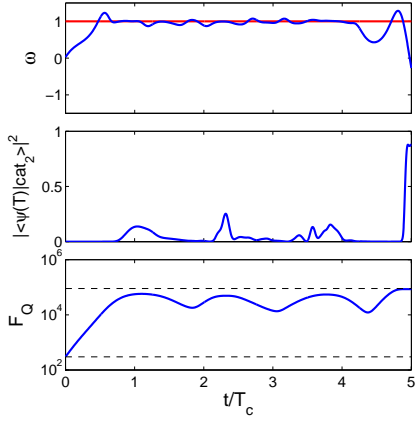}
\caption{\label{fig3} (Color online) Plot of the evolution of the
projection of the state $|\psi(t)\rangle$ onto the target state
$|\textrm{Cat}_1\rangle$ (middle), of the control field $\omega$ (top)
and of the Fisher information $F_Q$ (bottom). The control duration is $5\times T_c$. In the top panel,
the horizontal solid line is the solution of Ref.~\cite{Micheli}, which is taken
as a trial field of the algorithm.}
\end{figure}

The Hamiltonian (\ref{eq:ham_dopo_mapping}) of the
system is written as
\begin{equation}
\hat{H}=\hat{H}_0+\omega \hat{H}_1,
\end{equation}
where $\hat{H}_0=\chi \hat{J}_z^2$, $\hat{H}_1=\frac{\chi
N}{2}\hat{J}_x$ and $\omega$ is now the new control parameter. In
the following computations, we choose as parameters of the quantum
system $\chi=1$ and $N=300$.
The parameters of the algorithm are respectively taken to be
($\lambda=10^{-6}$, $\eta_1=\eta_2=10^{3}$) for $T=T_c$ and
($\lambda=5\times 10^{-4}$, $\eta_1=\eta_2=2$) for the two other
cases. The different computations are presented in Fig. \ref{fig3}
and \ref{fig5} for two durations $T=5\times T_c$ and $T=10\times
T_c$. We consider that the initial field of the algorithm is the
solution of Ref.~\cite{Micheli}, i.e. a constant field
$\omega(t)=1$ in the interval $[0,T]$. Due to the structure of the
control problem, note that the algorithm does not converge towards
an efficient solution for any initial condition. Taking as initial
field the solution $\omega(t)$ corresponding to higher values of
the bound in the geometrical protocol would not change
significatively the result because of the proximity of this
solution with the one of Ref.~\cite{Micheli}. 2000 iterations are
used in the first example, 1000 in the two others.

\begin{center}
\begin{figure}
\hspace{-1cm}
\begin{minipage}{.98\columnwidth}
\includegraphics[scale=0.15]{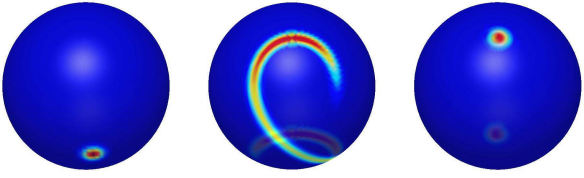}
\caption{\label{fig4} (Color online) Plot of the projections on the Bloch sphere, i.e. the Husimi function $Q(\theta,\phi) = |\langle \theta, \phi |\psi\rangle|^2$,
of the initial state $|\pi/2,\pi\rangle$ (left), the final states
with the geometric solution where $m=1$ and $T=T_{c}$ (middle) and with the monotonic
algorithm (right) where $T=10\times T_{c}$.}
\end{minipage}
\end{figure}
\end{center}

\begin{figure}[htbp]
\includegraphics[scale=0.4]{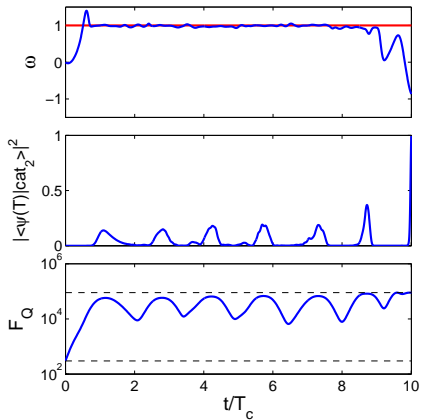}
\caption{\label{fig5} (Color online) Same as Fig. \ref{fig3} but for $T=10\times
T_{min}$.}
\end{figure}

\begin{figure}[htbp]
\includegraphics[scale=0.4]{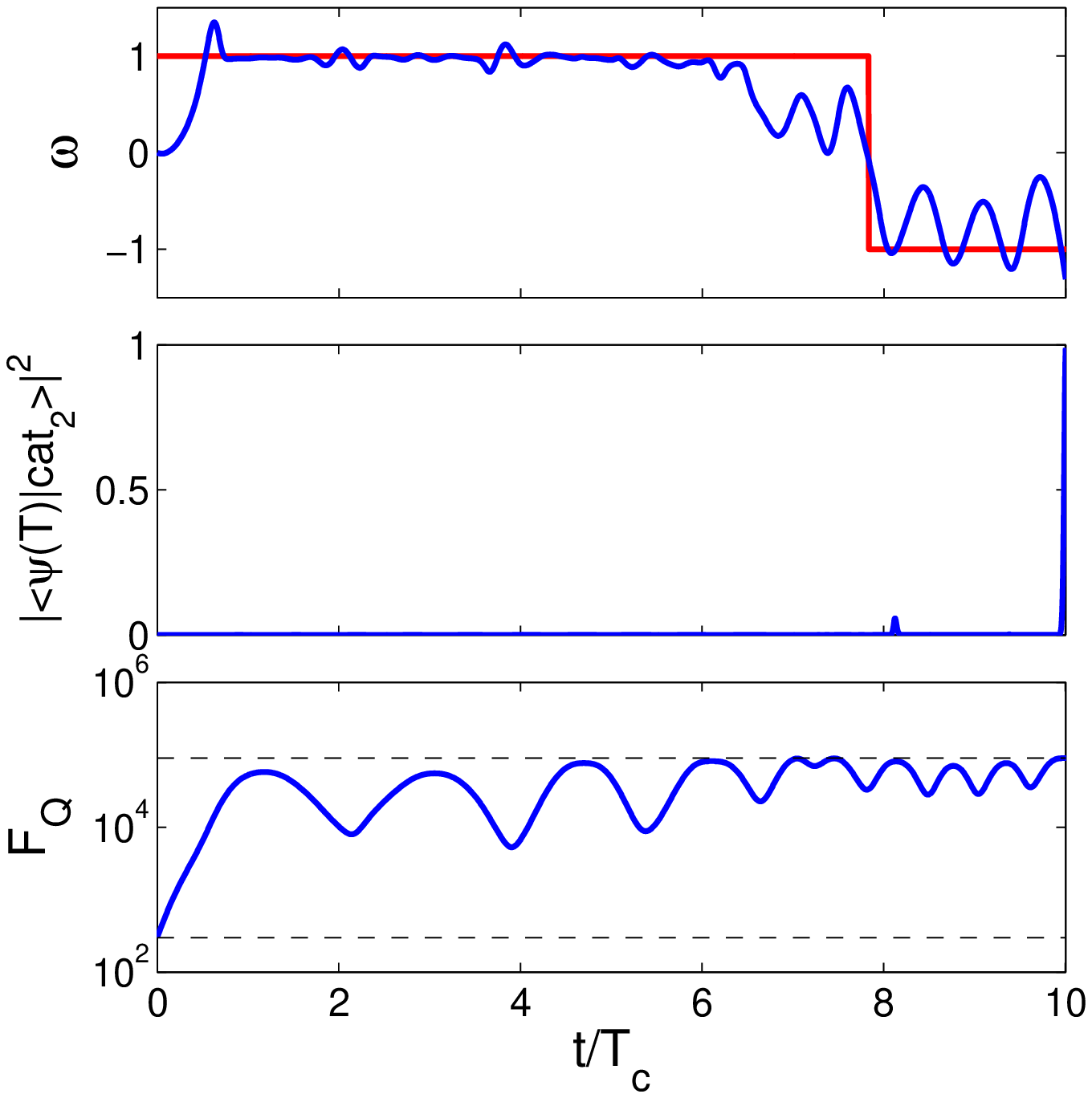}
\caption{\label{fig6} (Color online) Same as Fig. \ref{fig5} but for the target state $|\textrm{Cat}_2\rangle$.}
\end{figure}

We obtain that for the case  $T=5\times T_c$ and $T=10\times T_c$
the target macroscopic superposition can be created  with a very
high fidelity, respectively larger than 0.88 and 0.98,
corresponding to $F_Q/N^2 = 0.951$ and $F_Q/N^2 = 0.997$. Note
that for $T=T_c$, a projection of 0.2548 is reached, which is
better than the result provided in Ref.~\cite{Micheli}, but at the
price of a more complicated solution. This result was expected in
the sense that the solution by Micheli {\it et al.} has been
derived in the semi-classical limit without taking account the
quantum character of the problem. A visualization on the Bloch sphere of the state created is provided in the second panel of Fig. \ref{fig4}. The same computation has been
done for the target state $|\textrm{Cat}_2\rangle$ for $T=10\times
T_c$ and the parameters $\lambda=5\times 10^{-4}$ and
$\eta_1=\eta_2=2$. As initial condition for the algorithm, we
consider the bang-bang optimal solution with a total duration
increased by a factor of 10 (we keep constant the relative time of
the two bangs). We recall that a bang pulse is a field of maximum
intensity and a bang-bang sequence, the concatenation of two bang
pulses of different signs. In Fig. \ref{fig6}, note the bang-bang
structure of the final solution constructed by the algorithm. This
example shows that the algorithm has been guided towards a
particular mechanism by the semi-classical computation. The
different numerical results are listed in Table \ref{table2}. A
visualization on the Bloch sphere of the state created is provided
in the third panel of Fig. \ref{fig4}.

\begin{center}
\begin{table}[htbp]
\begin{tabular}{ccccccc}
            \hline
            &\multicolumn{3}{c}{NOON}&\multicolumn{3}{c}{PHASE}\\
            \hline
            n & 1 & 5 & 10 & 1 & 5 & 10\\
            \hline
            P & 0.255 & 0.880 & 0.994 & 0.245 & 0.903 & 0.989\\
            $F_Q/N^2$ & 0.650 & 0.951 & 0.997 & 0.632 & 0.946& 0.996 \\
            \hline
\end{tabular}
\caption{Same as Table \ref{table1} but for the quantum protocol. The parameter $n$ represents the ratio of the control duration over the time $T_c$ (see the text).\label{table2}}
\end{table}
\end{center}

In these control problems, we observe that a minimum time required to produce with efficiency a cat state (i.e. with a projection larger than 0.99) is of the order of $10\times T_c$. For the NOON state, we have analyzed the variation of this scaling factor with respect to the particle numbers $N$. Numerical simulations revealed that for $N$ ranging from 50 to 700 the target state is created with a very high precision for this duration, as can be seen in Table \ref{table-added} (to simplify the discussion, we only indicate the projection onto the target state and the Fisher information for a time of $10\times T_c$, even if we have tested other control durations).
This shows that $10\times T_c$ is a relevant time for this control problem, at least in the experimentally range of particle numbers considered.  For low values of $N$, it roughly corresponds to the minimum time where an efficient control field can be designed. For all the values of $N$ considered, the optimal solution is very close to a constant solution as shown in Fig. \ref{fig5} with $N=300$. Note that, when $N\geq 500$, we have numerically observed that lower times of the order of $7\times T_c$ can be chosen, but at the price of a more complicated structure for the field. This seems to be a general property of these quantum computations, i.e. the factor multiplying $T_c$ to consider to produce a superposition state decreases as $N$ increases. The determination of the exact dependence of this factor with the particle number is a hard task, since by construction the numerical algorithm does not allow to seek for the optimal time.

\begin{figure}
\hspace{-1cm}
\includegraphics[width=3.5in]{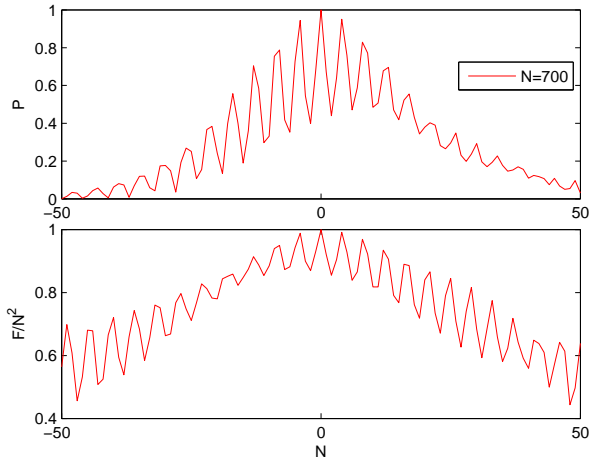}
\caption{\label{fig7} (Color online) (top) Plot of the projection
$P$ on the target state as a function of the variation of atom
number $\Delta~N$ for the optimal control field computed at
$N=700$ (see the text). The control duration is taken as $10\times
T_c$. (bottom) Same results but for the Fisher information.}
\end{figure}

\begin{center}
\begin{table}[htbp]
\begin{tabular}{cccc}
\hline
\hline
$N$ & $P$ & $F/N^2$ & $\chi t=10\times T_c(N)$\\
\hline
50   & 0.9995 & 0.9997 & 1.1976\\
100  & 0.9995 & 0.9996 & 0.6634\\
200  & 0.9997 & 0.9998 & 0.3688\\
300  & 0.9992 & 0.9997 & 0.2594\\
500  & 0.9998 & 0.9999 & 0.1659\\
700  & 0.9996 & 0.9999 & 0.1233\\
\hline
\hline
\end{tabular}
\caption{Numerical results of the optimal control computations for different atom number $N$. The control duration ($\chi t=10\times T_c$), the projection ($P$) and the Fisher information ($F_Q/N^2$) are given for different values of $N$.\label{table-added}}
\end{table}
\end{center}
In a final step, we have analyzed the robustness of the optimal
control field as a function of $N$, i.e. against the atom number
fluctuations. We choose as reference control field the one
corresponding to $N=700$. We have determined for $N$ ranging from
650 to 750 the projection and the Fisher information obtained at
$\chi t=10\times T_c$ with this control field. This point is
illustrated for the NOON state in Fig. \ref{fig7}. We observe that
the efficiency of the process quickly decreases when the particle
number changes. For instance, the average projection is already of
the order of 0.5 for a difference of 10 particles. This behavior
is not surprising in optimal control where the optimal solutions
are known to be very efficient (even for very complicated
problems), but are known to lack of robustness. A method based on
the simultaneous control of different systems can be used to
improve this latter property (see, e.g., Ref. \cite{skinner}).
This approach, which can be very time consuming in terms of
numerical computations, requires in addition to consider longer
optimal solutions than the ones used in this paper. This means
that the limit time of $10\times T_c$ is no more valid for
designing an optimal solution robust with respect to large
variations of $N$. Due to the complexity of this method, such a
computation goes beyond the scope of this article and has not been
undertaken.
\section{Experimental implementation of the protocol} \label{sec-exp}

We now analyze to which extend the fields corresponding  to the solutions found could be
implemented in real experiments with the present state of the art
technology. Let us discuss first the amplitude of the control.
>From Fig. \ref{fig5}, we see that the solution for the field
$\omega$, leading to the formation of a NOON state with a very
high fidelity in a time $10 \times T_c$, is roughly comprised
between $-1$ and $1.5$. Having in mind the internal BJJ setup (and
in particular the experiments of
Ref.~\cite{Oberthaler10,Gross_thesis}), typical experimental
bounds on the parameter $\Omega$ are $0 <  | \Omega  | < 2 \pi
\cdot 2$ kHz, and a typical value for $\chi$ is $\chi \approx 2
\pi \cdot 0.13$Hz. Fixing such a value for the interactions
translates into $0 < | \omega |< 102.6$, which largely comprises
the values required by our solution. Incidentally, we remark that
even the geometric solutions presented in Section \ref{sec4-1}
would be implementable, the largest value of $\omega$ considered
there being $\omega = 100$.
Furthermore, the control field can be switched fast compared with
the other time scales of the experiments. Note that it would be
also possible to include in our protocol some spectral constraints
on the control field \cite{mono}. Hence, in ideal conditions it
would be possible to implement our control protocol.

However, in realistic conditions the experiments are affected by the presence of noise, which induces dissipation and decoherence. The main sources of noise in the BJJ system are represented by particle losses~\cite{Sinatra} and phase noise~\cite{Ferrini_10}. This latter is due to stochastic fluctuations of the energies of the two modes of the BJJ. Other sources of noise are discussed in Refs.~\cite{Anglin,Witthaut,Moore06}. A quantitative discussion of the effect of decoherence on the creation of superposition states would require to treat on the same footing in our control protocol the unitary dynamics leading to the formation of the cat state and the decoherence sources. This point goes beyond the scope of this work, but represents an interesting perspective, which may be useful for the experimental quest of a macroscopic superposition. For a qualitative discussion, we can take the order of magnitudes of the decoherence rates of a process in which the macroscopic superposition is taken as initial state. We assume that this state evolves under the effect of noise only. In this case, we may distinguish the dissipation time $\tau_{\text{diss}}$, which characterizes the relaxation of the components of the superposition, from the decoherence time $\tau_{\text{dec}}$, i.e. the time at which the coherences of the superposition are washed out \cite{Orszag,Braun}. In general these two time scales are well separated. An exception is given by the decoherence induced on phase cat states by phase noise, for which the decoherence rate does not depend on the total particle number~\cite{Ferrini_10} (this is not the case for the NOON cat, for which $\tau_{\text{dec}}^{(-1)} \propto N$ under phase noise).
Losses processes can be one-body (due to scattering with impurities), two-body (spin-relaxation losses, due to collisions of two atoms which change their spin state, freeing a high kinetic energy and ejecting them out of the trap), or even three-body (when three atoms collide, two of them form a molecule, ejecting the third out of the trap \cite{Jack}).
The decoherence times are respectively $\tau_{\text{dec}}^{(-1)} \propto N$ for one body-losses, $\tau_{\text{dec}}^{(-1)} \propto N^2$ for two-body losses and $\tau_{\text{dec}}^{(-1)} \propto N^3$ for three body losses (see e.g. Ref.\cite{Jack} for the three body case).
Hence, if one-body losses and/or phase noise are affecting the system, since the time of formation of the cat states within the control protocol is of the order of $1/N$ and the decoherence rate grows linearly with the atom number, it is not obvious to determine the most convenient atom number. 
An analysis close to this spirit in the context of the creation of squeezed states in the presence of particle losses has been carried out by Li {\it et al} in Ref.~\cite{Sinatra}. If two and three-body losses are non negligible, since the decoherence rate grows respectively with $N^2$ and $N^3$ the experiments should be limited in practice to small atom numbers.

Let us consider again the order of magnitudes of the experiment in Ref.~\cite{Oberthaler10,Gross_thesis}, where the initial particle number is $N \approx 400$ (we have used $N = 300$ in our simulations).
The loss rate depends on the interaction strength, and is
typically of $\tau_{\text{diss}} \approx 0.1$s for $\chi \approx 2 \pi \times 0.13$ Hz \cite{Gross_thesis}. This corresponds to loosing about $15 \%$ of the atoms in about $20$ ms. Let us suppose that one-body losses are the dominant noise process. The time for the efficient generation of a NOON state is $10\times T_c$, resulting in $10 \times \chi T_c = 0.259$, i.e. $T_c = 0.317$ s, which is already of the same order of the dissipation time $\tau_{\text{diss}}$. Furthermore, the decoherence rate would be in this case $\tau_{\text{dec}} \sim \tau_{\text{diss}}/300 \sim 0.33$ms, which is a thousand time smaller than the time  $10\times T_c$. Therefore, in order for our protocol to be implementable with the experimental parameters \cite{Oberthaler10,Gross_thesis}, the dissipation rate should be lowered by a factor $1000$. In this framework, an analysis of the optimum atom number to use in the presence of decoherence could yield useful indications.

\section{Conclusions} \label{sec5}
In this paper we have designed different control fields by using
geometric and numerical optimal control techniques in order to create quantum superpositions in a bosonic Josephson junction. We have generalized the solution proposed by Micheli {\it et al.} in Ref.~\cite{Micheli} for the semi-classical model. 
The minimum time $T_{min}$ of formation of macroscopic
superpositions via our geometric control protocol has been shown
to be of the order of $1/N$, improving the result of
\cite{Micheli} by a numerical constant (see Appendix \ref{ana} for
the detailed calculation), with a fidelity of the same order. This
also shows the efficiency of the solution by Micheli {\it et al.},
which is very close to the time-optimal solution in the
semi-classical framework.

An almost perfect fidelity can be reached in  longer times of the
order of $10\times T_{min}$, with a more complicated field
solution. We have checked that this duration is sufficient for
creating very good superpositions for a large range of particle
number $N$, the fidelity further improving at increasing $N$. Note
that such a duration can still lead to a significative speed-up
with respect to the protocol based on the quenched dynamics of the
BJJ \cite{noi}, where the characteristic time of formation of a
two-component superposition of phase states is $\pi/(2 \chi)$,
independent on $N$ \cite{footnote-other-cats}.

Decoherence effects may change significatively the state crated
with the designed control field, and a careful analysis of the
control sequence to be used in the presence of noise stems as an
important extension of the present work.
\appendix

\section{Optimal control methods}\label{sec3}
\subsection{Geometric approach}\label{secgeo}
We present in this section the geometric optimal theory which
allows us to solve the optimal control problem in the
semi-classical limit. In this work, we consider a system belonging to a general
class of optimal control problems on a two-dimensional manifold
(here the Bloch sphere) with a single control field. Powerful
mathematical tools that we briefly describe below have been
developed in this case \cite{boscainbook}.

We consider a system governed by a differential equation of the
form
$$
\dot{\vec{x}}=\vec{F}(\vec{x})+u\vec{G}(\vec{x}),
$$
where $\vec{x}$ is a two-dimensional state vector, $\vec{F}$ and
$\vec{G}$ two vector fields and $u$ a control field satisfying the
constraint $|u|\leq m$, where $m$ is a bound on the control.
Starting from a point $\vec{x}_0$, the goal of the control is to
reach in minimum time a target $\vec{x}_f$. We assume that a
solution exists, i.e. that the target belongs to the accessibility
set of the initial point $\vec{x}_0$. We solve this time-optimal
control problem by applying the Pontryagin Maximum Principle (PMP),
which is formulated from the pseudo-Hamiltonian
\begin{equation}\label{eqham}
H=\vec{p}\cdot (\vec{F}+u\vec{G}),
\end{equation}
where $\vec{p}$ is the adjoint state. The Hamiltonian trajectories which are solutions of
\begin{equation}
\dot{\vec{x}}=\frac{\partial H}{\partial
\vec{p}}(\vec{x},\vec{p},v),~\dot{\vec{p}}=-\frac{\partial
H}{\partial \vec{x}}(\vec{x},\vec{p},v),
\end{equation}
where the control field $v$ is determined from the maximization
condition
\begin{equation}\label{eqmax}
H(\vec{x},\vec{p},v)=\max_{|u|\leq m} H(\vec{x},\vec{p},u),
\end{equation}
with $H\geq 0$,
are candidates to be optimal. Such trajectories are called
extremals in the control community. Other geometric tools have to
be used (see \cite{boscainbook} for a recent overview on this
point) to select among the extremals, the ones which are
effectively optimal.

The maximization condition (\ref{eqmax}) can be solved from the switching
function $\Phi$ given by
\begin{equation}
\Phi(t)=\vec{p}\cdot \vec{G}.
\end{equation}
Using the maximization equation, one deduces that an extremal curve $(x(t),p(t))$ is composed of a sequence of arcs $\gamma_+$,
$\gamma_-$ and $\gamma_s$. Regular or bang arcs $\gamma_\pm$ are obtained
when $\textrm{sign} [\Phi]=\pm 1$. In this case, the control field
is given by $u=m\times\textrm{sign} [\Phi]$. When the function
$\Phi$ takes a zero value and when this zero is isolated, the
field switches from $\pm m$ to $\mp m$. We encounter a singular
situation and singular arcs if $\Phi$ vanishes on an interval
$[t_0,t_1]$. In this case, the control field cannot be directly
determined from the maximization condition, but from the
fact that $\dot{\Phi}=\ddot{\Phi}=\cdots= 0$ on $[t_0,t_1]$. A
straightforward computation leads to
\begin{eqnarray*}
\dot{\Phi}=\vec{p}\cdot [\vec{G},\vec{F}]
\end{eqnarray*}
where $[\vec{G},\vec{F}]$ is the commutator of two vector fields
defined by
\begin{equation*}
[\vec{G},\vec{F}]=\vec{\nabla} \vec{G}\cdot \vec{F}-\vec{G}\cdot
\vec{\nabla} \vec{F},
\end{equation*}
$\vec{\nabla}\vec{G}$ being the Jacobian matrix of the vector
$\vec{G}$. One then deduces that the system $\Phi=\dot{\Phi}=0$
admits a non-trivial $\vec{p}$ solution on the singular set $S$
defined by
\begin{equation}
S=\{\vec{x};\textrm{det}(\vec{G},[\vec{G},\vec{F}])(\vec{x})=0\},
\end{equation}
where the vector fields $\vec{G}$ and $[\vec{G},\vec{F}]$ are
parallel. The singular arcs $\gamma_s$ are located in the set $S$.
The singular control field $u_s$ can be computed from the second
derivative of $\Phi$
\begin{equation*}
\ddot{\Phi}=\vec{p}\cdot
[\vec{G},[\vec{G},\vec{F}]]+u_s\vec{p}\cdot
[\vec{F},[\vec{G},\vec{F}]]=0,
\end{equation*}
which leads to
\begin{equation}
u_s=-\frac{\vec{p}\cdot [\vec{G},[\vec{G},\vec{F}]]}{\vec{p}\cdot
[\vec{F},[\vec{G},\vec{F}]]}.
\end{equation}
Note also that the optimal solution can follow the singular lines
only if the control field is admissible, that is if
$|u_s(\vec{x})|\leq m$.
\subsection{Monotonically convergent algorithms}\label{monoto}
Monotonically convergent algorithms are a standard approach to
solve the optimality equations in quantum mechanics. Such
algorithms have been described in detail elsewhere \cite{tannorbook,mono}. Here, we only
summarize the main theoretical aspects to apply these methods
\cite{mono}.

We respectively denote by $|\phi_0\rangle$ and $|\phi_f\rangle$
the initial and target states of the control problem and we assume
that the Hamiltonian of the system can be written as
\begin{equation}
\hat{H}=\hat{H}_0+u(t)\hat{H}_1,
\end{equation}
where $u(t)$ is the control field. We consider the following cost
functional which penalizes the energy of the control field
\begin{equation}
J=|\langle \phi_f|\psi(T)\rangle |^2-\lambda \int_0^T u(t)^2dt
\end{equation}
where $\lambda$ is a positive parameter. The value of $\lambda$
expresses the relative weight between the projection onto the
target state and the energy of the field. The control duration $T$
is assumed to be fixed here. In order to satisfy the constraint
that the state $|\psi(t)\rangle$ is solution of the Schr\"odinger
equation, we introduce the augmented cost $\bar{J}$
\begin{eqnarray*}
\bar{J}=& &|\langle \phi_f|\psi(T)\rangle |^2-\lambda \int_0^T
u(t)^2 dt\\
& & -2\times \textrm{Im} [\langle \psi(T)|\phi_f\rangle \int_0^T \langle
\chi(t)|(i\frac{\partial}{\partial t}-\hat{H})|\psi(t)\rangle dt],
\end{eqnarray*}
where $|\chi(t)\rangle$ is the adjoint state and $\textrm{Im}$
denotes the imaginary part of a complex number. Using the fact
that the variations of $\bar{J}$ with respect to
$|\psi(t)\rangle$, $|\chi(t)\rangle$ and $u$ are equal to 0, one
arrives at a differential system of the form
\begin{eqnarray*}
& &(i\frac{\partial}{\partial t}-\hat{H}(t))|\psi(t)\rangle =0\\
& & (i\frac{\partial}{\partial t}-\hat{H}(t))|\chi(t)\rangle =0,
\end{eqnarray*}
with the initial conditions $|\psi(0)\rangle=|\phi_0\rangle$ and
$|\chi(T)\rangle=|\phi_f\rangle$. The control field $u$ is given
by
\begin{equation}
u(t)=-2\times \textrm{Im}[\langle \psi(T)|\phi_f\rangle\langle
\chi(t)|\hat{H}_1|\psi(t)].
\end{equation}
This set of coupled equations is solved by using monotonic
convergent algorithms but other approaches such as gradient
algorithms could be used \cite{grape}. We assume here that the
iteration is initiated from a field $u_0(t)$. At step $k$ of the
algorithm, the system is described by a triplet $(|\psi_k\rangle,
|\chi_{k-1}\rangle, u_k)$ corresponding to the cost $J_k$
\begin{equation}
J_k=|\langle \phi_f|\psi_k(T)\rangle |^2-\lambda \int_0^T
u_k(t)^2.
\end{equation}
Between two iterations, the variation $\Delta J$ is given by
\begin{eqnarray*}
\Delta J=J_{k+1}-J_k &=& |\langle \psi_{k+1}(T)|\phi_f\rangle
|^2-|\langle \psi_{k}(T)|\phi_f\rangle |^2 \\
& & -\lambda \int_0^T (u_{k+1}(t)^2-u_k(t)^2).
\end{eqnarray*}
Introducing $A_{k,k+1}=-2\textrm{Im}[\langle \psi_{k+1}|\chi_k\rangle
\langle \chi_k |H_1|\psi_{k+1}\rangle ]$, one arrives at
\begin{eqnarray*}
\Delta J &=& \int_0^T dt \lqu
(\tilde{u}_k-u_{k+1})A_{k,k+1}+(u_k-\tilde{u}_k)A_{k,k} + \right. \nn \\
& & \left. \lambda (\tilde{u}_k^2-\tilde{u}_k^2+u_{k+1}^2-u_k^2) \rqu,
\end{eqnarray*}
which can be written as
\begin{equation}
\Delta J=\int_0^T dt (P_1(t)+P_2(t)),
\end{equation}
where
\begin{eqnarray*}
& & P_1=-(u_{k+1}-\tilde{u}_k)[\lambda(u_{k+1}+\tilde{u}_k)+A_{k,k+1}],\\
& & P_2=-(\tilde{u}_k-u_k)[\lambda(u_{k}+\tilde{u}_k)+A_{k,k}].
\end{eqnarray*}
To ensure the monotonic behavior of the algorithm, we choose the
fields $u_{k+1}$ and $\tilde{u}_k$ such as the integrands $P_1$
and $P_2$ are positive. More precisely, we first
determine $\tilde{u}_k$ from $u_k$ such that $P_2$ is
positive and then we determine $u_{k+1}$ from $\tilde{u}_k$ such
that $P_1$ is positive. This means that we define $\tilde{u}_k$
from $P_2$ as
\begin{equation}\label{eqeta2}
\tilde{u}_k-u_k=-\eta_2 [\lambda (u_k+\tilde{u}_k)+A_{k,k}]
\end{equation}
where $\eta_2$ is a positive constant. The field $\tilde{u}_k$ can
finally be expressed as
\begin{equation}\label{eqfield1}
\tilde{u}_k=\frac{1-\lambda \eta_2}{1+\lambda
\eta_2}u_k-\frac{\eta_2}{1+\lambda \eta_2} A_{k,k}.
\end{equation}
The same work can be done for the field $u_{k+1}$ and $P_1$. In
this case, we get
\begin{equation}\label{eqeta2}
u_{k+1}-\tilde{u}_k=-\eta_1 [\lambda
(u_{k+1}+\tilde{u}_k)+A_{k,k+1}],
\end{equation}
$\eta_1$ being also a positive constant, and
\begin{equation}\label{eqfield2}
u_{k+1}=\frac{1-\lambda \eta_1}{1+\lambda
\eta_1}\tilde{u}_k-\frac{\eta_1}{1+\lambda \eta_1} A_{k,k+1}.
\end{equation}
Note that this formulation is equivalent to the one introduced in
\cite{mono}.

The structure of the algorithm can be summarized as follows.
\begin{enumerate}
\item Backward propagation of the adjoint state $|\chi_k(t)\rangle$
with the field $\tilde{u}_k$ and the initial condition
$|\chi_k(T)\rangle=|\phi_f\rangle$. The field $\tilde{u}_k$ is
determined at each time through the formula (\ref{eqfield1}).
\item Forward propagation of the state $|\psi_{k+1}(t)\rangle$ with
initial condition $|\psi_{k+1}(t)\rangle=|\phi_0\rangle$ and the
field $u_{k+1}$ which is computed at the same time by the formula
(\ref{eqfield2}).
\end{enumerate}
Using this procedure, it is then straightforward to show that the
algorithm remains monotonic, that is $J_{k+1}\geq J_k$.
Numerically, the algorithm is stopped when its convergence is
better than a given threshold $\varepsilon$, i.e.
$|J_{k+1}-J_k|\leq \varepsilon$.


\section{Measure of the coherence of the macroscopic superposition}\label{sec_tests}

After deriving the control field $\omega(t)$, we may quantify the coherence of the macroscopic superposition state which has been created by means of the control protocol. In this section, we present some quantities which will be used to this scope. Note that, in the case of geometric control, such methods require the computation of the quantum evolution under the field $\omega(t)$.


\subsection{Fidelity}

The first method consists in computing the projection (or fidelity) of the final state onto the cat states introduced in
Eq.~(\ref{cat1},\ref{cat2})
\be
\label{eq:fidelity}
P_{1,2} = |\langle \textrm{Cat}_{1,2} | \psi(t) \rangle |^2,
\ee
to be evaluated at the time of formation of the superposition. However, such a quantity does not describe the character of the state, and in particular its shape and its coherence.

\subsection{Angular momentum eigenvalues probability distributions}
\label{sse:p_n}

In order to represent the macroscopic superpositions created with our protocol, we can use the probability distribution of the eigenstates of angular momentum operators in various directions \cite{Ferris, Ferrini_2}. These probability distributions carry different informations when considering different directions and states. We consider as a first example the ideal NOON state given in Eq.~(\ref{cat1}). In this case, the probability distribution of the eigenvalues associated to the Fock states, defined as the projection $P(n) = |\langle n | \textrm{Cat}_{1} \rangle |^2$ where $|n\rangle$ satisfies Eq.~(\ref{eq:fock_states}), is given by
\be
P(n) = \fr{1}{2} \lt \delta_{n,N/2} + \delta_{n,-N/2} \rt.
\ee
This distribution displays two peaks corresponding to the two components of the superposition at $n = \pm N/2$ (see the bottom panel of  Fig.~\ref{fig1}). The analogous $\hat{J}_x$-eigenvalues distribution for the same state, i.e. $P_{x}(n_x) = |\langle n_x |\textrm{Cat}_{1} \rangle|^2$ where $|n_x\rangle$ satisfies $\hat{J}_x | n_x \rangle = n_x | n_x \rangle$, corresponds to the profile of the NOON state when projected on the $x$ axis. A straightforward calculation using Eqs.~(\ref{eq-coherent_state}) and (\ref{eq:fock_states}) leads to
\be
\label{px}
P_{x}(n_x) =
\begin{cases}
\fr{1}{2^{N-1}} {N \choose \fr{N}{2} + n}
& \text{if $n$ is even}
\\
 0
&   \text{if $n$ is odd}.
\end{cases}
\ee
This corresponds to a binomial envelop centered in $n = 0$, with interference fringes having an unit spacing (see the bottom panel of Fig. \ref{fig2}) \cite{Ferris, Ferrini_2}. Note that the same distribution for an incoherent mixture of the same two coherent states, such as $\hat{\rho}_1=\frac{1}{2}(|\theta = 0, \phi = 0\rangle \langle \theta = 0, \phi = 0| + |\theta = \pi, \phi = 0\rangle \langle \theta = \pi, \phi = 0|)$, would display no fringes.

Since the phase cat state (\ref{cat2}) is the rotation of the NOON state lying along the $x$ axis, the preceding arguments have to be reversed for such a state. In particular, its probability distribution $P_{x}(n_x)$ displays two peaks at $n_x = \pm N/2$, while the Fock states probability distribution $P(n)$ can be useful to visualize the coherences.

\subsection{Quantum Fisher information}

Additional insights about the coherence of the superposition state, i.e about its off-diagonal correlations, can be gained from the computation of the optimum quantum Fisher information $F_Q = 4\times \text{max}_{\vec{n}} (\Delta J_{\vec{n}})^2$ \cite{Caves94, Smerzi09, Smerzi10}. Such a parameter was historically introduced to quantify the usefulness of a quantum state for interferometry \cite{Caves94}, and it has been demonstrated that the condition $F_Q > N$ provides a sufficient condition for the state to be entangled \cite{Smerzi09}. In particular, for the macroscopic superposition states $|\textrm{Cat}_{1,2}\rangle$ given in Eqs.~(\ref{cat1},\ref{cat2}) $F_Q = N^2$, while $F_Q$ scales linearly with the particle number for an incoherent mixture of coherent states, such as the mixture $\hat{\rho}_1$ introduced in Sec.~\ref{sse:p_n}.
In this work, this parameter will be used as an indicator of the presence of off-diagonal correlations in the state produced within our control protocol.

\section{Analytic expressions of the control durations}\label{ana}
We derive in this section the expressions of the durations both
for the solution of Ref.~\cite{Micheli} and the time-optimal trajectory. By
definition of the classical control problem from the quantum one,
we assume that the distance between the point of coordinates
$O=(\pi/2,\pi)$ and the initial point $I$ of the dynamics is
$\sigma= \sqrt{N}$, this point belonging to the separatrix for
the solution of Ref.~\cite{Micheli} and to the singular set for the optimal one.
There exist two different natural ways to define this distance,
one corresponding to the length of the segment $OI$, i.e. the
euclidian distance, the other to the length of the geodesic on the
sphere from $O$ to $I$. Denoting by $(\theta,\phi)$ the coordinates of $I$, one gets in
the first case that
$$
\frac{\sigma}{\sqrt{N}}=\sqrt{2(1-\cos\phi\sin\theta)},
$$
while in the second case
$$
\sin^2(\frac{\sigma^2}{N})=\sin^2\theta\sin^2\phi+\cos^2\theta.
$$
We assume in the computation of $T_{min}$ that there is no bound
on the control field, so that the time of travel along the bang
arc can be neglected. In this case, $T_{min}$ is the time taken to
travel from $I$ to the point of coordinates
$(\theta=3\pi/4,\phi=\pi/2)$. The dynamics is described by the
following set of equations,
\begin{eqnarray*}
& & \dot{\phi}=\frac{\chi
N}{2}(-2\cos\theta-\omega\cot\theta\cos\phi) \\
& & \omega=\sin\theta\cos\phi \\
& & \cos^2\theta=\sin^2\phi\sin^2\theta
\end{eqnarray*}
where the second and third equations correspond respectively to
the definitions of the singular control field and of the singular
set. Straightforward computations lead to
\begin{equation}
T_{min}=\int_{\phi}^{\pi/2}\frac{dx}{\frac{\chi N}{2}\sqrt{\sin^2
x(1+\sin^2 x)}},
\end{equation}
where $\phi$ is determined from the expression of $\sigma$. A similar argument can be used to compute $T_c$ for the solution of Ref.~\cite{Micheli} along the separatrix. In this case, the dynamics is
governed by the equations
\begin{eqnarray*}
& & \dot{\phi}=\frac{\chi
N}{2}(-2\cos\theta-\omega\cot\theta\cos\phi) \\
& & \omega=1 \\
& & \cos\phi=\sin\theta
\end{eqnarray*}
where $\cos\phi=\sin\theta$ is the equation of the separatrix
\cite{Micheli}. This gives
\begin{equation}
T_c=\int_{\phi}^{\pi/2}\frac{2dx}{\chi N \sin x}
\end{equation}
which can be simplified into
\begin{equation}
T_c=-\frac{2}{\chi N}\ln (\sqrt{2N}-\sqrt{2N-1})
\end{equation}
for the euclidian distance and into
\begin{equation}
T_c=-\frac{2}{\chi N}\ln
(\frac{1-\sqrt{\cos\sigma}}{\sqrt{1-\cos\sigma}})
\end{equation}
for the geodesic on the sphere. In the limit $N\to +\infty$, we
recover in the two cases the expression (\ref{eq:time_cat}) given in \cite{Micheli}.
\begin{figure}[htbp]
\includegraphics[scale=0.4]{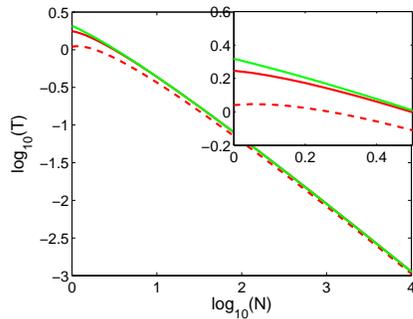}
\caption{\label{fig8} Evolution of the control duration in the
semi-classical model (see the text) as a function of the number of
particles $N$. The red (dark gray) and green (light gray) solid
lines represent respectively the exact and the approximated time
for the Micheli solution. The dashed line corresponds to the
optimal solution with $m=1\times 10^6$. The small insert is a zoom
near $N=0$ of the plot.}
\end{figure}
The difference between the two control durations is illustrated in Fig. \ref{fig7}. One observes that this difference decreases as the number of particles $N$ becomes larger, which shows the efficiency of the Micheli solution in terms of optimality for the creation of quantum superpositions.\\ \\
\textbf{ACKNOWLEDGMENTS} We thank A. Minguzzi, F.W.J. Hekking, D. Spehner and C. Gross for useful discussions.\\



\begin{thebibliography}{33}
\bibitem{notazioni} In this work we will follow the notations of Refs.~\cite{Smerzi97,Chr_preprint_bifurcation} instead of the ones in Ref.~\cite{Micheli}.
\bibitem{footnote-other-cats} By the quenched dynamics of the BJJ superpositions of $N$ phase states (instead of two) are formed in a time inversely proportional to the square root of the number of particles $N$ \cite{noi-last}.
\end{thebibliography}
\end{document}